# Giant Gain Enhancement in Photonic Crystals with a Degenerate Band Edge

Mohamed A. K. Othman[1], Farshad Yazdi[1], Alex Figotin[2] and Filippo Capolino[1]

[1]Department of Electrical Engineering and Computer Science, University of California, Irvine, Irvine, CA,.92697 USA

[2]Department of Mathematics, University of California, Irvine, Irvine, CA, 92697 USA

{mothman, fyazdi, afigotin, f.capolino}@uci.edu

We propose a new approach leading to giant gain enhancement. It is based on unconventional slow wave resonance associated to a degenerate band edge (DBE) in the dispersion diagram for a special class of photonic crystals supporting two modes at each frequency. We show that the gain enhancement in a Fabry-Pérot cavity (FPC) when operating at the DBE is several orders of magnitude stronger when compared to a cavity of the same length made of a standard photonic crystal with a regular band edge (RBE). The giant gain condition is explained by a significant increase in the photon lifetime and in the local density of states. We have demonstrated the existence of DBE operated special cavities that provide for superior gain conditions for solid-state lasers, quantum cascade lasers, traveling wave tubes, and distributed solid state amplifiers. We also report the possibility to achieve low-threshold lasing in FPC with DBE compared to RBE-based lasers.

## I. INTRODUCTION

Light confinement using either mirrors or Bragg reflectors provides for high quality ($Q$)-factor in Fabry-Pérot cavity (FPC) resonators and enhanced optical field intensity. Such cavities are commonly used for laser applications and spectroscopy. An important class of high $Q$-factor structures is formed by slow-wave resonators based on the regular band edge (RBE) of the wavenumber-frequency dispersion diagram relative to photonic crystals, whose simplest architecture is a periodic stack of dielectric layers, with one dimensional periodicity [1–3]. More elaborate designs of nanocavities adopted Silicon heterostructures [4], liquid crystals [5] technologies and demonstrated improved $Q$-factor compared to previously reported designs. The use of photonic crystals resulted in enhanced amplification properties for low-threshold lasing [2,6], enhanced directional-wave propagation through magneto-optical effects [7–9], nonlinear optics [10] and quantum processing [11].

Pursuing better performing photonic crystal cavities is essential to further advancement of photonic technology [12–15], and photonic integrated circuits [16–18] in particular. These advancement established a basis for a novel class of solar cell architecture with enhanced absorption [19–21], and other thin film applications [22], along with superior atomic interaction with strongly localized photons [23], and unconventional spontaneous emission dynamics [24,25]. Slow light in photonic crystals is yet another fundamental utility that can tailor the electromagnetic response and achieve superior performance through dispersion engineering [26–28].

Figotin and Vitebsky in [29–33], proposed FPC resonators made of *unconventional* photonic crystals composed by anisotropic dielectric layers. Those FPC resonators exhibit sharper transmission peaks, higher $Q$-factors, and better general performance in a vicinity of the photonic band edge frequency compare to conventional photonic crystal FPCs of the same size made of isotropic layers. The related field enhancement properties in those unconventional structures can be attributed to the degenerate band edge (DBE) conditions.

This special DBE condition produces some four electromagnetic modes (EM) at the DBE frequency; that phenomenon does not occur in regular photonic crystals, i.e., conventional photonic crystals exhibiting an RBE providing a single EM mode operation. Consequently, it is important to acknowledge that the resonance characteristics in DBE cavities studied in this paper are fundamentally different from those in standard band-gap cavities [2,3,34,35]. Significant differences between DBE and RBE based FPCs are highlighted in Sec. II.

The principal result of this paper that the DBE condition based on resonance properties discussed in [29–33] lead to giant power gain when an active material is integrated into the periodic structure. The gain material in our studies is modeled by complex refractive index (see section 2.5 in [36] or section 8.2 in [39]) and we show that the achieved power gain in an FPC with a DBE can be of several orders of magnitude higher than the same produced by a comparable photonic crystal with RBE and the same active medium. In the both cases slow-light is involved, but we demonstrate here that the DBE condition is far superior to the RBE condition for enhancing the power gain.

An example of giant gain in a periodic DBE structure is clearly shown by the peak in Fig. 1 (red curve). This periodic structure is made of three layers: two anisotropic and one isotropic, exhibiting a DBE as those in [30–32], and the amplification occurs when active material is introduced. The layered stack with three different colors in the inset of Fig. 1 shows a DBE-based structure. The method used to calculate the transmission power gain is detailed in section III. The layer that includes an active material, and thus intrinsic gain, can be realized by either layers of quantum wells [40], Raman scattering mechanisms [41] or layers doped with active materials like Erbium [42] for example. A possible implementation of anisotropic layers can be carried out utilizing liquid crystals, for instance, possessing tunable anisotropic properties [5,43].

We would like to point out that the conceptual findings reported in this paper not only apply to stacks of anisotropic layers, but also to various other periodic guiding structures that exhibit the DBE condition, such as optical coupled waveguides/nanowires systems made with silicon-on-insulator technology embedded with gratings, as those in [44–47]. An attractive application for such amplification scheme is quantum cascade lasers which are superior compared to the ones based on isotropic type of photonic crystals as in [35], with a promise of very high quantum efficiency.

## II. BACKGROUND AND METHODS

Our main goal is to show a possibility of giant amplification in a gain medium that can exceed the levels occurring in conventional structures. Such a gain medium is provided by a periodic structure supporting a DBE as we demonstrate here for a specific example based on some preliminary ideas expressed in [48]. To understand origins of the giant gain let us recall some basic properties of DBE in photonic crystals. In Fig. 2(a-b) we show the dispersion diagram (that relates frequency to the complex Bloch wavenumber $k$) for modes associated with a specific periodic structure used in our demonstrations. (Note that in [29–33] only the real part of these Bloch wavenumbers was shown). The Bloch modes may exhibit an RBE condition at a certain frequency $\omega_g$. When two or more modes are allowed to propagate at the same frequency a crystal may also exhibit either a DBE at a frequency $\omega_d$ or the split band edge (SBE) conditions. A stationary inflection point (SIP) also can occur in magnetic photonic crystals. Waves with $(k,\omega)$ in the vicinity of those stationary points of the dispersion diagram exhibit almost vanishing group velocity as a consequence of multiple reflections. These waves experience an effective increase in the optical path. Indeed, the increase in the group index at an RBE condition was a basis for various studies suggesting possible gain enhancement in FPCs with RBE (see, for example, [34,49] and references therein), while the concept of SIP was investigated theoretically for unidirectional lasers [9].

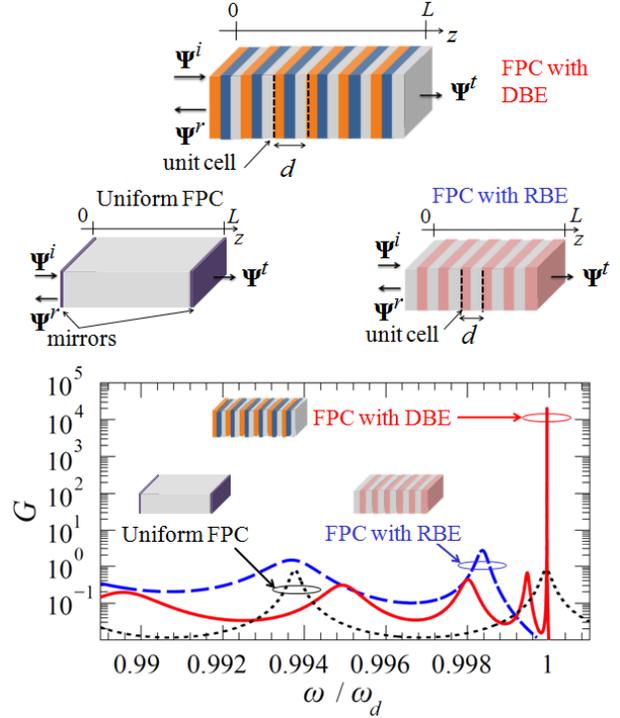

FIG. 1. Total transmission power gain $G$ for three cavities: uniform FPC (dotted-black); FPC formed by a periodic structure of isotropic layers that exhibit an RBE at $\omega_d$ (dashed-blue); and a FPC formed by a special photonic crystal, like a stack of misaligned anisotropic (birefringent) layers, that exhibit a DBE at $\omega_d$ (solid-red). In the top geometries, the unit cell boundaries are marked by dashed black lines. Active material is embedded in only the isotropic (gray) layers, having $n'' = -2.2 \times 10^{-4}$.

The wavenumber-frequency dispersion relation for the propagating mode can be approximated by the asymptotic expression $\Delta\omega \propto (\Delta k)^2$ near an RBE, and by $\Delta\omega \propto (\Delta k)^4$ near a DBE, where the increments are with respect to the stationary condition. We will show that this implies a gigantic increase in the density of states (electromagnetic modes) and the group index near a DBE compared to what happens near an RBE. This leads to also to $Q$-factors that are orders of magnitude higher than the same for their RBE counterparts, namely $Q \sim N^5$ [30], with $N$ being the number of unit cells of the dielectric stack.

The $Q$-factor improvement in FPCs with DBE can be understood by observing a larger amount of average electromagnetic energy stored inside the resonator, compared to the energy leaking outside (when dissipation due to material losses is negligible compared to the power leaking out). In other words, the cavity effective mode volume [50] shrinks significantly close to the DBE condition, leading to high levels of field enhancement. We will quantitatively relate the enhancement of the $Q$-factor to the giant gain through concept of photon lifetime, in section III-B, and also to the local density of states in section V. Such intriguing features of DBE structures, which we study here, can facilitate a gigantic boost in power gain, in addition to efficient manipulation of the lifetime of quantum emitters, the Purcell factors as well as nonlinear effects in DBE-based structures.

## III. GIANT GAIN AND THE WIGNER TIME INCREASE

We show in this section that the gain in an FPC operated at DBE is enhanced by several orders compared to one with an RBE condition (Fig. 1). This giant gain enhancement can be explained by a significant increase of the Wigner time compared to that for other cavities. To see that let us consider three kinds of FPCs. The first one is a conventional FPC formed by an isotropic and homogenous dielectric material bounded by two highly reflective mirrors, referred to as uniform FPC, as shown in Fig. 1. The second one is made by stacking isotropic dielectric bi-layers (referred to as FPC with RBE and shown in Fig. 1). The third one is made by stacking anisotropic tri-layers (referred to FPC with DBE) as depicted in Fig. 1, with two anisotropic layers of the same thickness $0.31d$, and an isotropic one of thickness $0.38d$, with parameters given in Appendix A. The two periodic structures have the same unit cell thickness $d$, and the FPC length is $L = Nd$. We assume that the FPC with RBE is constructed from alternating lossless dielectric layers with real refractive indices $n_1 = 3.2$ and $n_2 = 1.5$ with equal thickness designed to have a band edge at $\omega_g = \omega_d$ for the sake of comparison. The uniform FPC has a refractive index of $n = 2.25$ (see Appendix A for the case when active materials are present in these FPCs). Note that the selection of boundaries for the unit cell is depicted in the inset of Fig. 1 by dashed lines. Inside the anisotropic layers, the $x$ and $y$ polarizations are coupled therefore we elaborate on the electromagnetic field representation inside those layers. The guided time-harmonic electromagnetic fields (varying as $e^{-i\omega t}$) along the $z$-direction associated to these *two* coupled waves are described by a *four* dimensional state vector $\mathbf{\Psi}(z) = \begin{bmatrix} E_x & E_y & H_x & H_y \end{bmatrix}^T$, which satisfies Maxwell's equations and boundary conditions at the interfaces between layers, and evolves along the $z$-direction [30,32]. The transformation across a unit cell of length $d$ is described by using a 4×4 transfer matrix $\underline{\mathbf{T}}(z+d,z)$ that relates the field between two points $z$ and $z+d$ as $\mathbf{\Psi}(z+d) = \underline{\mathbf{T}}(z+d,z)\mathbf{\Psi}(z)$. Such transfer matrix is found by cascading the transfer matrices of the constitutive layers $\underline{\mathbf{T}}_m$ as $\underline{\mathbf{T}}(z+d,z) = \prod_{m=1}^{M} \underline{\mathbf{T}}_m$, with $M$ being the number of layers per unit cell (for example $M = 3$ for the DBE structure analyzed here, and $M = 2$ for the RBE structure), and $\underline{\mathbf{T}}_m = \underline{\mathbf{T}}_m(z+d_m, z+d_{m-1})$ where $d_m$ is the thickness of the $m$th layer, and $d_0 = 0$. Details of calculation of such transfer matrix are presented in Appendix A. By solving the eigensystem $\underline{\mathbf{T}}(z+d,z)\boldsymbol{\Psi}(z) = \lambda \boldsymbol{\Psi}(z)$, one obtains the eigenvalues $\lambda$ related to the Bloch wavenumber $k$ as $\lambda = e^{ikd}$ defining the modal propagation. The DBE condition is equivalent to the requirement for $\underline{\mathbf{T}}(z+d,z)$ to be similar to a Jordan block [30,31].

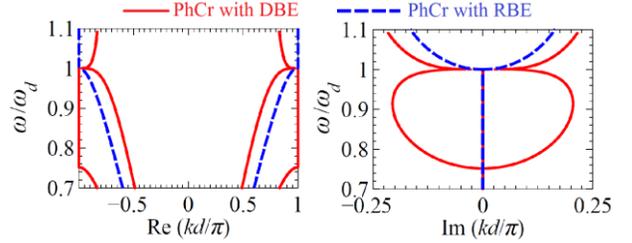

FIG. 2. Dispersion relations $k-\omega$, showing the real and imaginary parts of the Bloch wavenumber for the two photonic crystals (PhCr) in the inset of Fig. 1: one that exhibits only RBE at $\omega_d$ (dashed) and another that exhibits DBE at $\omega_d$ (solid).

The Brillouin zone dispersion relation for the periodic structures whose unit cell defines the periodic type of FPCs is depicted in Fig. 2, showing the real and imaginary parts of the eigenmode Bloch wavenumber $k$. For symmetry reasons, both $k$ and $-k$ are modal solutions. Accordingly, the photonic crystal with pairs of isotropic layers supports only one propagating mode in the each direction below the RBE (for instance at $\omega = 0.75\omega_g$ with $\omega_g = \omega_d$). Instead, the structure with anisotropic layers is able to support two modes simultaneously propagating in the each direction below the RBE (for instance at $\omega = 0.75\omega_d$) and only one propagating mode (the other is evanescent) in the frequency range above the RBE and below the DBE at $\omega = \omega_d$. Considering the two structures in Fig. 1, just below $\omega_d$ there exists only one propagating mode, in each direction, for the FPC with RBE and the FPC with DBE, as seen in Fig. 2. This propagating mode in the $+z-$direction (pertaining to the $+k$ solution in the unbounded periodic structure) has a unique state vector denoted by $\boldsymbol{\Psi}_1$, whose electric field components are described by the vector $\mathbf{E}_1 = \begin{bmatrix} E_{x1} & E_{y1} \end{bmatrix}^T$.

For a finite stack of $N$ unit cells as in the inset of Fig. 1 we report the exact total transmission power gain in Fig. 1 calculated as $G = P_{\text{out}} / P_{\text{inc}}$ where $P_{\text{inc}}$ and $P_{\text{out}}$ are the time-average incident and transmitted power densities, respectively. We consider a transverse electromagnetic (TEM) plane wave excitation with an electric field vector $\mathbf{E}^i$ that is polarized along the vector $\mathbf{E}_1$. The electric fields transmitted through the FPC comprise a vector $\mathbf{E}^t$, and the incident and transmitted power densities are calculated as $\frac{1}{2}\|\mathbf{E}^i\|^2/\eta$ and $\frac{1}{2}\|\mathbf{E}^t\|^2/\eta$, respectively, with $\eta$ being the wave impedance of the surrounding material, assumed to be vacuum, and $\|.\|$ represents the norm of the vector. In this example we take $n'' = -2.2 \times 10^{-4}$ describing the isotropic active material with intrinsic gain, represented by the grey layers in the geometries in Fig. 1(keeping the real part of their corresponding refractive indices the same as above, see Appendix A). Such value used for $n''$ can be realized by layers of quantum wells [38,40], or by Erbium doping [42], for example, and $n''$ is related to the complex electric susceptibility of the active medium (details of such description are found in [36–40]). Moreover we assume the FP resonance frequency of the cavity falls within the active material emission band.

For the DBE case we assume the active layers to be embedded only in the isotropic region. Therefore the active material filling factor $f$ is 100% for the uniform FPC, $f = 50\%$ for the FPC with an RBE, and $f = 38\%$ for the FPC with a DBE (38% is the filling factor of the isotropic layers in the FPC with DBE as explained in Appendix A), and both FPCs have $N=32$ unit cells. Despite the smallest active filling fraction, the structure with DBE is the one that provides giant amplification, up to four orders of magnitude larger than the FPC with RBE, as shown in Fig. 1where we compare the total amplification provided by the structure with an RBE structure and a uniform FPC. The reasons for such giant enhancement in gain is quantified in the following section.

### A. Transmission characteristics

We calculate here the transmission coefficient $T$ for the same stack of $N$ unit cells shown in Fig.1. This coefficient is polarization dependent on the impinging TEM plane wave due to the anisotropy of some layers in the FPC with DBE. We select the same incident polarization used to calculate the gain in Fig. 1., i.e., matched to the polarization state $\mathbf{E}^i$ of the mode supported by the infinite structure. We define the transmission coefficient for FPC of length $L$ as the projection of the transmitted electric field vector $\mathbf{E}^t$ on $\mathbf{E}^i$, i.e., $T = \left(\mathbf{E}^t, \mathbf{E}^i\right) / \|\mathbf{E}^i\|$ where the parenthesis denote the inner product of those two vectors in complex normed space, with $\|\mathbf{E}^i\|$ as the norm.

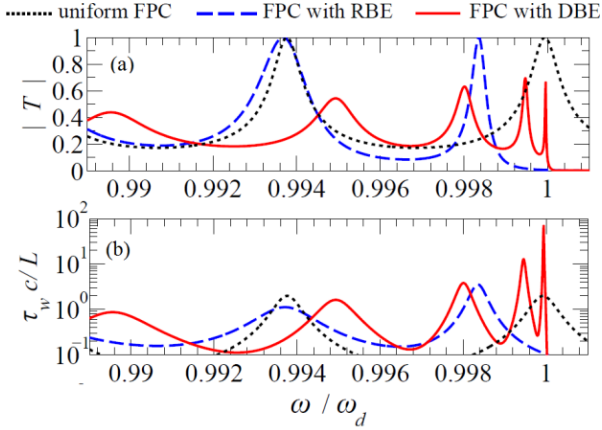

FIG. 3. (a) Magnitude of the transmission coefficient, (b) Wigner time for three types of FPCs: uniform FPC (dotted), FPCs formed by a periodic structures that exhibit either an RBE (dashed) or a DBE (solid) at $\omega_d$.

We observe in Fig. 3(a) a very narrow (high $Q$-factor) transmission peak for the FPC with a DBE, with $N = 32$ unit cells. This occurs at an angular frequency very close and lower than to the DBE angular frequency $\omega_d$, because strong reflection occurs at $\omega_d$ and for higher frequencies due to an existing bandgap [10,11]. The transmitted spectrum is densely-packed with resonances, for the FPC with DBE near the DBE frequency. By neglecting the phase shift introduced by the interfaces, the wavenumber of the DBE resonance mode for large number of unit cells $N$ is estimated by $k_{r,d} \approx \pi/d - \pi/(Nd)$, and the corresponding resonance frequency $\omega_{r,d}$ can be calculated as $\omega_{r,d} \approx \omega_d - b(\pi/(Nd))^4$ [30,31] where $b$ is a problem specific constant. For the RBE case we have a similar approximation of the resonance frequency, with $1/N^2$ instead of $1/N^4$ [3,34,51]. In addition, FPCs in Fig. 1, with RBE and DBE, respectively, exhibit a high quality factor resonance without using mirrors whereas the uniform FPC necessitates mirrors to reach the same quality factors of an RBE FPC cavity. Such mirrors, designed here with ~95% power reflectivity can be realized by, for example, large stacks of Bragg reflectors.

### B. Wigner time and photon lifetime

Transmission properties considered in previous section are the first indicators of a possibility of enhancing gain in FPCs with a DBE. Estimation of the photon lifetime inside the FPC is yet another pass to origins of the gain enhancement. Before assessing the photon lifetime, we notice that the Wigner time of FPC (also referred to as tunneling time or group delay) is another important quantity that can be easily obtained for our structure. The Wigner time is defined as

$$\tau_w(\omega) = d\varphi(\omega)/d\omega, \qquad (1)$$

where $\varphi(\omega)$ is the total transmission phase shift accumulated across the FPC of length $L$, i.e., the phase of the $T$ coefficient. As explained in [25–27], this quantity is a measure of the effective group velocity of photons passing through the FPC. Fig. 3(b) shows the normalized Wigner time for the three FPCs under study, showing more than an order of magnitude increase for the FPC with a DBE, near $\omega_d$. The other two types of FPCs have similar Wigner time since the resonances of those have similar $Q$ factor for the chosen cases. Although Wigner time calculated here is based on the phase of the transmission coefficient defined previously, we have observed (not shown here) that at the DBE $\tau_w$ becomes almost polarization independent owing to the rapid variation of the phase, $\varphi(\omega)$ at the DBE resonance.

The Wigner time is required to calculate the effective optical path length $L_{\text{eff}}$, that is the length required for photons to traverse the cavity, calculated as $L_{\text{eff}} = n_{\text{eff}} L$, where $n_{\text{eff}}$ is the effective group index of the cavity. Accordingly, the effective group index is given as $n_{\text{eff}} = c\tau_w/L$ where $c$ is the speed of light in vacuum, and $\tau_w$ is the Wigner time of the cavity before introducing materials with intrinsic gain. The effective length, $L_{\text{eff}} = c\tau_w$ dramatically increases near a DBE as a consequence of the Wigner time increase. When an active material, described by complex-valued refractive index $n = n' + in''$ with $n'' < 0$, is introduced in the FPC, the photon *single pass* power gain coefficient $g_0$ of the composite medium is given by $g_0 = \gamma L f$, where $f$ is the active material filling factor in the cavity and $\gamma$ is the per-unit-length material intrinsic gain coefficient. The latter is related in turn to the imaginary part of the wavenumber inside that active material as $\gamma = -2k_0 n''$ with $k_0 = \omega/c$ [52]. We define then an effective gain coefficient $g_{\text{eff}}$ as the gain experienced by photons traveling through the FP cavity, that is

$$g_{\text{eff}} = \gamma L_{\text{eff}} f = g_0 L_{\text{eff}}/L = g_0 n_{\text{eff}} = \gamma f c \tau_w. \qquad (2)$$

This reveals that the effective gain coefficient of an FPC is the product of the intrinsic material gain coefficient and the FPC Wigner time. The above formula is valid provided that $\gamma$ is sufficiently small not to deteriorate the modal properties of the DBE resonance, as discussed in the following section. Notice also that the transmission power gain in general is polarization dependent, and there exists specific polarizations maximizing the gain. However, it is the Wigner time that is a decisive factor for the giant power gain enhancement with far more significant impact on the power gain than the incident polarization. The total power gain $G$ at a DBE resonance can be estimated using the effective gain coefficient, as follows. We denote by $G_{\mathrm{approx}}$ the following approximation to the total power gain

$$G_{\mathrm{approx}} = |T|^2 e^{g_{\mathrm{eff}}}, \qquad (3)$$

where the transmission coefficient $T$ is evaluated at resonance for the structure without gain (Fig. 3(a)). This is a simple way to approximate the small-signal amplifier transmission power gain from the transmission and the gain coefficients. Note that the scaling with the transmission coefficient is necessary here to obtain an estimate of the total power transmission gain $G_{\mathrm{approx}}$, since $|T| \neq 1$ at a DBE resonance as can be seen in Fig. 3. In section V we quantify and compare an estimate of the total power gain as in formula (3) and the power gain shown in Fig. 1. When conceiving structures with large effective gain coefficient one has to keep in mind that its dependence on the cavity and material properties as described by formula (2) cannot hold if the resonance is largely perturbed as discussed in the next section.

It is also instructive to compare the Wigner time with the photon lifetime $\tau_c$, which is defined as time it takes the stored energy to decay to $1/e$ fraction of its original value due to dissipation. Such a quantity is frequently used in the laser physics to express the resonator finesse [37,39,52] as well as to provide a good estimation of the Winger time, as explained in many different situations [51,53–57]. Although lifetime and group delay are different concepts, they are intimately related. Indeed, for a uniform lossless FPC at resonance we have an exact identity $\tau_c = \tau_w$ [55,57], which means that stored energy lifetime in such cavity identically equals to the time it takes the wave packet to traverse the whole FPC [57], but this identity is not exactly satisfied in general. The importance of $\tau_c$ relies in its explicit link to the Q-factor via $Q = \omega \tau_c$ [39,52], whereas there exist approximations of the $Q$ factor using the Wigner time [54]. The lifetime quantifies the evolution of energy with time inside the FPC. Near the DBE, $\tau_c$ is extremely large because the quality factor is enhanced [30,33] leading to eminently high levels of stored energy. This may have impacts on significantly lowering the lasing threshold condition in laser devices, as we point out in the following section.

## IV. CRITICAL GAIN AND THRESOLHD CONDITION

Since DBE is a precise mathematical condition, i.e., it only emerges when the transfer matrix **T** becomes similar to a Jordan Block. Consequently any perturbation (such as loading the cavity with active material with $n'' < 0$) can eventually deteriorate this condition. To put things into perspective, large single pass gain $g_0$ can deteriorate the slow-wave phenomena condition associated with the DBE (similarly for the RBE case shown in [49]) implying that large enhancement of photon lifetime will cease to occur. On the other hand, usually very small single pass gain $g_0$ does not result in a strong amplification. However, there exists a threshold for intrinsic gain, or starting oscillation condition, where lasing oscillations start to occur. Hence, we define a threshold gain coefficient $\gamma_{th} = -2k_0 n''_{th}$ as the minimum amount of intrinsic gain necessary and sufficient to maintain lasing in the cavity through the stimulated emission. In other words, it is the intrinsic gain such that the round trip losses via energy escaping from the cavity ends and internal losses are compensated exactly by the round trip gain introduced by the active material. That mathematical condition is found using the complex frequency poles loci of the transmission coefficient or transfer function $T$ [58]. Below threshold the FPC acts as a linear (small signal) amplifier as described earlier. We show here that in theory an FPC with a DBE is going to have a lower lasing threshold than any other comparable cavity due to the very high gain coefficient of that cavity. An example of threshold behavior will be shown in the section VI.

## V. GIANT ENHANCEMENT OF LOCAL DENSITY OF STATES AND GAIN

As a last step toward providing a comprehensive understanding of the modal properties of an FPC with DBE that contribute to giant gain, we investigate the local density of states (LDOS) inside the FPC with DBE. Interesting features of the spontaneous emission in particular is related to the LDOS. Near the band edge of a photonic crystal, the spontaneous emission characteristics of an atom or an "emitter" can be significantly altered under some conditions and requires cavity quantum electrodynamic treatment [24,50,59–61]. For the sake of simplicity, we use classical concepts that agree with more general quantum mechanical description of the interaction between atoms and photons in the weak coupling regime [50,60]. We also assume that an emitter has a line shape much broader than the DBE resonance spectral width. For that purpose we dedicate this section to investigate the LDOS in our novel cavity using classical electromagnetic description utilizing the Green's function [50,60,62,63] and show that such LDOS calculations provide a good estimate of the FPC gain calculated via transfer matrix approach. The emitter is modeled as an electric sheet current source with the current density of the form $\mathbf{J} = J_0 \hat{\mathbf{p}} \delta(z-z')$, where $\hat{\mathbf{p}}$ is a transverse unit vector with $\hat{\mathbf{z}} \cdot \hat{\mathbf{p}} = 0$, and consequently only transverse-to-z electromagnetic waves contribute to the LDOS in this case for consistency (see Appendix B for details). This one dimensional treatment is valid when all waves emitted have a wavevector in the z-direction (normal to the layers). It provides physical insights into the use of such cavity. In the more realistic case where an emitter radiates in all directions in the three dimensional space, precise spectral formulation of the emission can be carried out in the multi-layered environment as developed in [64,65]. We are interested in the average of the projected (or partial) LDOS $\rho(\hat{\mathbf{p}}, z)$ over all possible orientations $\hat{\mathbf{p}}$. Accordingly, the average z-dependent LDOS $\rho(z)$ in an open, lossless FPC is proportional to the trace of the 2×2 dyadic Green's function $\underline{\underline{\mathbf{G}}}(z,z)$ and given by [60,62,63,66]

$$\rho(z) = \langle \rho(\hat{\mathbf{p}}, z) \rangle_{\hat{\mathbf{p}}} = -\frac{k_0}{\pi c} \operatorname{Im}\left( \operatorname{Tr}\left[ \underline{\underline{\mathbf{G}}}(z,z) \right] \right), \qquad (4)$$

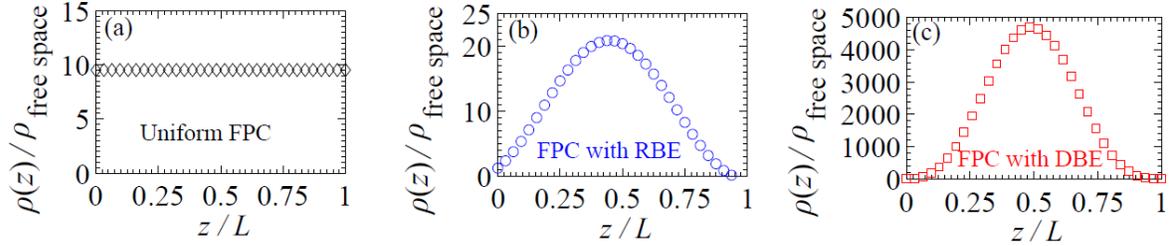

FIG. 4. Sampled local density of states (LDOS) normalized by the LDOS in free space inside (a) uniform FPC sampled at maxima locations, (b) FPC with RBE and (c) FPC with DBE sampled at points $z = md$, $m = 0,1,2,\ldots,N$ with $N = 32$.

where Im and Tr denote respectively the imaginary part and the trace, and $\langle \cdot \rangle_{\hat{\mathbf{p}}}$ denotes the average over all possible emitter polarization directions. We construct then 2×2 matrix (dyadic) Green's function $\underline{\underline{\mathbf{G}}}(z,z)$ that provides transverse electric field response to a Dirac delta sheet current polarized in $x$ and/or $y$ directions at $z$ = constant. The precise definition of $\underline{\underline{\mathbf{G}}}(z,z)$ and its properties are given in Appendix B. For the particular case where a DBE is occurring, we apply the transfer matrix formalism, analogously to what was done in [51,67,68] for a simpler photonic crystal made by a stack of isotropic layers as detailed in Appendix B. We report in Fig. 4(a-c) the LDOS normalized by the LDOS in free space for the three types of FPCs considered in this paper. For simplicity the LDOS is sampled once per unit cell in the two periodic structures (those with RBE and DBE) since it varies also within each unit cell. We observe that the LDOS for the FPC with DBE or with RBE peaks around the cavity center. Vice versa the uniform FPC has a uniform LDOS envelope across the length, when evaluated as sampling the LDOS at each maximum of the electric field and the envelope is independent on the resonant mode; only the number of maxima depends on the chosen resonance.

The largest LDOS enhancement occurs for the DBE case, and it is due to the unconventional DBE resonance inside the cavity that generates giant fields.

These giant fields are caused by a strong excitation of both propagating and evanescent modes in the FPC, as a necessity to satisfy the boundary condition at the FPC edges. The required field continuity at the boundary is obtained by destructive interference between intense propagating and the evanescent internal modes. At the center of the cavity the evanescent modes become negligible whereas the propagating modes of the giant intensity become dominant as explained in details in [30,31]. The LDOS enhancement up to two orders of magnitude is quite remarkable and can be related to the giant amplification regimes, as described next.

As it is shown in [25] the Wigner time $\tau_w$ is proportional to the spatial average of LDOS over the length of the cavity $\langle \rho(z) \rangle_L$, where the brackets denote a spatial average over the length $L$. In other words, we may write $\tau_w = K \langle \rho(z) \rangle_L$ where $K$ is a constant, that is a function of the dielectric contrast of the layers [51]. (A detailed analysis of how to determine constant $K$ is outside the scope of this paper, however it can be developed following the procedure detailed in [51]).

It is important to note that the effective gain coefficient $g_{\text{eff}}$ in (2) is in turn proportional to the average of the LDOS and we can also write the gain coefficient enhancement factor as proportional to the ratio between LDOS to that in free space

$$g_{\text{eff}} = g_0 \frac{c \tau_w}{L} = g_0 \frac{\langle \rho(z) \rangle_L}{\rho_{\text{free space}}} K, \qquad (5)$$

where $\rho_{\text{free space}} = 1/(\pi c)$. In our case we evaluate $K$ as $K = \langle \rho(z) \rangle_L / \tau_w$, which leads to $K = 0.045$ at the DBE resonance. Accordingly, we compare the total power transmission gain $G$ calculated in Fig. 1 at the DBE resonance, via transfer matrix method as ratio of transmitted and incident fields, and the one calculated by (3) with $g_{\text{eff}}$ estimated by either (2) or (4). We recall that the transmission coefficient $T$ in (3) is evaluated for the structure without gain (Fig. 3(a)). The total transmission gain $G$ calculated via transfer matrix method at the DBE resonance in Fig. 1 as field ratio at its peak is $G \cong 2 \times 10^4$, while using the latter approximate formula leads to a peak $G_{\text{approx}} \cong 2.015 \times 10^4$ when considering a single pass gain of $g_0 = 0.11$ based on the value of $n'' = -2.2 \times 10^{-4}$ that in turns results in $g_{\text{eff}} \cong 10.7$. This indicates an evident agreement between the exact gain and the estimation using the LDOS or the Wigner time in (5), and proves that the giant gain enhancement in intimately related to the giant boost in the Wigner time and LDOS.

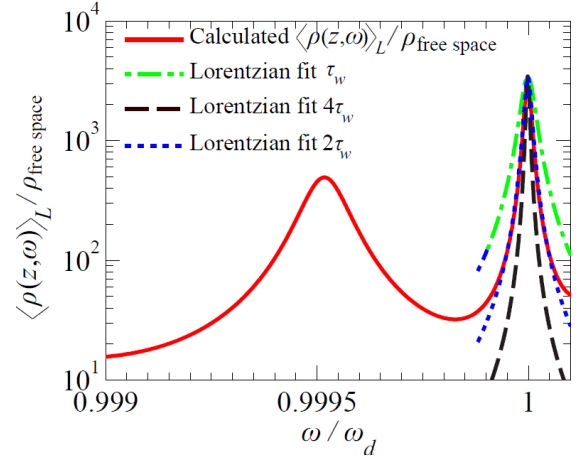

FIG. 5. Profile of frequency variation of the spatial average of LDOS in the FPC with DBE with $N = 32$ unit cells, near the DBE radian frequency $\omega_d$. We also plot a Lorentzian fitting with different values of the fitting parameters $h$.

To further discuss the dynamical features that may arise near the DBE (following a similar method discussed in [24,25,59,69,70] for the case of an RBE), we now consider the strongly varying spectral behavior of the LDOS near the band edge. We are interested only in the frequencies close to the DBE resonance peak, at $\omega_{r,d}$, close to $\omega_d$ since there the largest LDOS is observed. Since the DBE resonant spectral width is much narrower than the separation between other adjacent FPC resonant frequencies, we can approximate the spatial average of LDOS around $\omega_{r,d}$ by a Lorentzian profile (as typically done for high Q

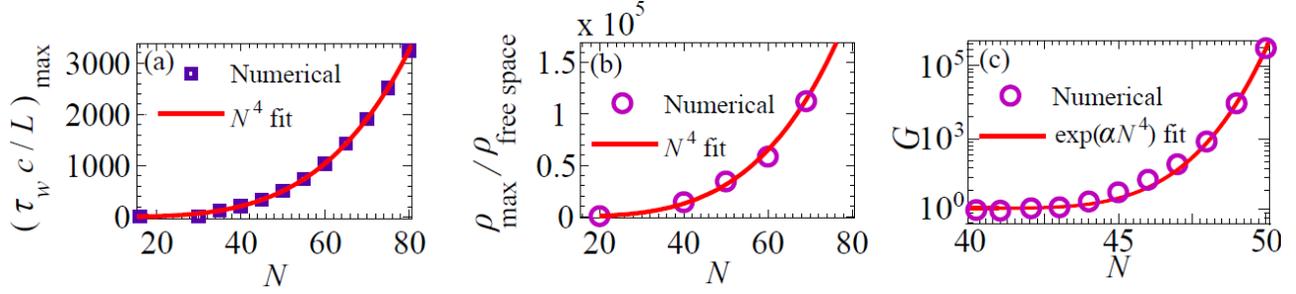

FIG.6. (a-b) Trend of the Wigner time and LDOS varying the number of unit cells $N$ showing an accurate fit to $N^4$. Plot (c) shows the scaling of the total transmission power gain $G$ versus $N$, with a fitting of $G \sim \exp(\alpha N^4)$.

resonators' LDOS [61,71,72]) given by

$$\langle \rho(z,\omega) \rangle_L \cong \frac{\rho_{r,d}}{1 + h^2(\omega - \omega_{r,d})^2}, \quad (6)$$

for $|\omega - \omega_{r,d}| \ll \omega_{r,d}$, where $h$ is a fitting constant related to the cavity $Q$ factor, $\rho_{r,d} \equiv \langle \rho(z) \rangle_L \big|_{\omega_{r,d}}$ is the peak spatial average LDOS evaluated at the resonance radian frequency $\omega_{r,d}$. In the limit where $Q \to \infty$, i.e., no photon dissipation, the LDOS diverges, corresponding to a closed resonator, i.e., $\langle \rho(z,\omega) \rangle_L \to \delta(\omega - \omega_{r,d})$ where $\delta(\omega - \omega_{r,d})$ is the Dirac-delta function. In Fig. 5 we plot the calculated spatial average of the LDOS varying as a function of frequency near the DBE. We also report the fitting using the formula (6) for a few values of the fitting parameter $h$. It is noted that the best fitting is provided by $h = 2\tau_w$, as expected because the $Q$ factor of the cavity $Q = \omega_{r,d}\tau_c$. Therefore this results shows that in our case $\tau_c \approx \tau_w$ and $h = 2Q/\omega_{r,d} = 2\tau_w$ for the best fitting. This proves that $\tau_c \approx \tau_w$ is a valid assumption for FPCs with DBE with high $Q$.

There characteristics can be readily incorporated in calculating spontaneous emission, by following the same procedure developed in [24,60] and by incorporating the band edge condition $\omega - \omega_d = a(k - k_d)^4$ and the LDOS Lorentzian profile in (6) in calculating the decay response of atoms in cavity, and exploring other interesting features such as strong coupling regimes [69,70,73] that can be benefit from high Q cavities with large LDOS.

## VI. SCALING WITH LENGTH

Finally, we investigate here the gigantic scaling of the Wigner time and the LDOS with the cavity length, as well as the lasing threshold of the FPC, at the DBE resonant radiant frequency $\omega_{r,d}$. First of all, increasing the number of unit cells, or equivalently the cavity size $L$, will increase the effective gain coefficient $g_{\text{eff}}$ (and hence the total power gain $G$), quality factor, and LDOS for an FPC. However, for a *uniform cavity* those parameters can be easily determined analytically, where, for instance, we can determine that the quality factor is *linearly* proportional to the length of the FPC. For an FPC with RBE, the Wigner time and the LDOS are varying as $N^2$ as calculated using asymptotic analysis [3], and the $Q$-factor scales as $N^3$, as already observed in [3,33].

We have verified here for the first time that an FPC with DBE provide a giant enhancement of the Wigner time and LDOS, that both scale as $N^4$. This is evidently a superior enhancement compared to cavities based on the RBE previously studies. In Fig. 2 we have already shown that a FPC filled with some active material with intrinsic gain exhibits a giant gain. Based on the presented analysis, we can also show that the total transmission gain $G$, is proportional to the exponential of the effective gain coefficient as in (5) and scales as $G \sim \beta \exp(\alpha N^4)$ where $\alpha$ and $\beta$ are fitting constant. Indeed, the effective gain coefficient scales as $g_{\text{eff}} \sim N^4$ following the scaling of the Wigner time and the LDOS. For a constant material gain per unit cell in the cavity $n''$, by increasing the number of unit cells the total gain $G$ substantially increases as shown in Fig. 6(c) for $n'' = -10^{-5}$, and we find the exponential fitting parameters $\alpha \cong 5.2 \times 10^{-7}$ and $\beta \cong 4.7 \times 10^3$. Note that in Fig. 6(c), below $N = 40$, there is no gain ($G < 1$) since the small material gain is not sufficient to overcome the losses (energy escaping out the cavity ends). However, for large number of cells, the same material gain is sufficient to compensate losses and cause a large gain enhancement, and with proper choice of parameters may also lead to oscillations.

For that purpose, and to demonstrate a potential application for FPCs with DBE, we are also interested in obtaining the threshold conditions for starting oscillation in such cavity as described in section IV. In Fig. 7 we plot the intrinsic normalized lasing threshold $\gamma_{th} / \gamma_0$ as function of the DBE and RBE length $N$. The calculation is based on the transfer matrix method as described in Sec. IV. Note that for the sake of comparison, one may recall that the threshold gain coefficient in a uniform cavity with mirrors (uniform FPC) is inversely proportional to the length of the cavity [37,60] as $1/L$. However we see that the DBE structure threshold is asymptotically fitted to $1/N^5$ while for the RBE cavity it is fitted to $1/N^3$. For instance, FPC with DBE lasing threshold condition is $\gamma_{th}$ = 48.33 [m$^{-1}$] whereas for FPCs with RBE the threshold is $1.84 \times 10^3$ [m$^{-1}$] for $N$ =32. These trends are determined by numeric fitting the exact data in Fig. 4(d-f) since analytical methods have not been developed yet for a photonic crystal with a DBE. Based on these observations, it can be inferred that a FPC with DBE of smaller length can suffice to achieve even higher gain $G$ and lower threshold lasing condition compared to a much larger FPC with RBE or filled with uniform material.

## CONCLUSION AND REMARKS

We have shown that the degenerate band edge (DBE) condition pertaining to some special photonic crystals that can support a degeneracy with some four EM modes at a specific frequency, such as an anisotropic stack of dielectric layers, can be utilized to strongly enhance the gain in Fabry-Pérot cavities. This property is strictly related to the degeneracy condition at the edge of the Brillouin zone. We have demonstrated that due to extremely large Wigner time and very high local density of states for DBE structures there is consequent giant amplification which is of orders of magnitude larger than the same obtained for uniform FPCs or for FPCs with RBE. We have also found lower lasing thresholds for DBE based cavities compared to RBE counterparts. One example of a potential application for extremely high gain could be in Erbium doped fiber amplifiers (EDFA) and quantum cascade lasers. High $Q$-factor and giant gain scaling offered by DBE structures can be utilized also in $Q$-switching and mode locking lasers. For instance, by varying the misalignment angles of the anisotropic layers [74], the dispersion diagram of the periodic structure can be readily altered and hence the $Q$ factor [75] and gain can be significantly switched. Other tuning conditions can be developed with other guiding structures. These mechanisms have potential in laser applications and pulse compression optical components.

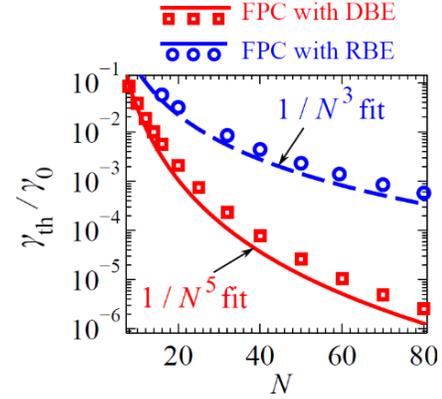

FIG. 7. Calculated lasing threshold for oscillation for both DBE and RBE based FPCs (symbols) and their respective fitting to $1/N^5$ and $1/N^3$. A DBE based FPC exhibits a much lower threshold.

DBE-based giant gain enhancement can be useful also for enhancing the amplification in microwave traveling wave tubes (TWT). Indeed, our preliminary results indicate that interaction of a very low energy electron beam with electromagnetic modes in a corrugated waveguide with DBE results in giant amplification. Analogous mechanisms can also be used in backward-wave-oscillator sources. In addition to that, radio frequency and microwave solid state distributed amplifiers can also benefit from using active components distributed along a microwave transmission line exhibiting the DBE.

## ACKNOWLEDGEMENT

This research was supported by AFOSR MURI Grant FA9550-12-1-0489 administered through the University of New Mexico. Authors acknowledge also support from AFOSR Grant FA9550-15-1-0280.

## APPENDIX A: TRANSFER MATRICES OF ANISOTROPIC LAYERS

Plane wave propagation along the *z* direction in a dielectric described by a permittivity tensor follows the simple relation derived from Maxwell equations

$$\frac{\partial}{\partial z}\mathbf{\Psi}(z) = ik_0 \underline{\mathbf{M}}(z)\mathbf{\Psi}(z), \quad (A1)$$

where $\underline{\mathbf{M}}(z)$ is a 4×4 matrix operator [30] and $\mathbf{\Psi}(z)$ is the state-vector as in Sec. III. In the case where the dielectric is described by a symmetric relative permittivity tensor in the Cartesian coordinates reference as [30,31]

$$\begin{pmatrix} \underline{\underline{\boldsymbol{\varepsilon}}}_t(z) & 0 \\ 0 & \varepsilon_{zz} \end{pmatrix}; \quad \underline{\underline{\boldsymbol{\varepsilon}}}_t(z) = \begin{pmatrix} \varepsilon_{xx}(z) & \varepsilon_{xy}(z) \\ \varepsilon_{xy}(z) & \varepsilon_{yy}(z) \end{pmatrix}, \quad (A2)$$

an expression of the operator $\underline{\mathbf{M}}$ is given by [10,11]

$$\underline{\mathbf{M}}(z) = \begin{bmatrix} 0 & 0 & 0 & \eta \\ 0 & 0 & -\eta & 0 \\ -\varepsilon_{xy}(z)/\eta & -\varepsilon_{yy}(z)/\eta & 0 & 0 \\ \varepsilon_{xx}(z)/\eta & \varepsilon_{xy}(z)/\eta & 0 & 0 \end{bmatrix}. \quad (A3)$$

The transfer matrix described in Sec. III is then constructed from the solution of the Cauchy problem in (A1) with the appropriate boundary conditions. In short, the transfer matrix of relative to an individual layer operator $\underline{\mathbf{M}}_m$ is calculated as $\underline{\mathbf{T}}_m = \exp(i\underline{\mathbf{M}}_m d_m)$ [30]. The anisotropic layers considered in this papers have parameters taken from [31], $\varepsilon_{xx} = 16.31$, $\varepsilon_{yy} = 5.79$ and $\varepsilon_{zz} = 1$, however the physical concepts described in this paper are general and can be applied also to several other structures supporting two coupled modes. The first anisotropic layer in the unit cell in Fig. 1 has an orientation $\alpha$ of its optical axes with respect to the reference system and has $\varepsilon_{xy} = 5.26$, whereas the second layer has an orientation $-\alpha$, hence it has $\varepsilon_{xy} = -5.26$. This corresponds to a misalignment $2\alpha$ between the two adjacent anisotropic layers of 45 degrees in the *x-y* plane, with $\alpha = 22.5$ degrees. The third layer in the unit cell in Fig. 1 (made of three layers) is isotropic with $n^2 = \varepsilon_{xx} = \varepsilon_{yy} = \varepsilon_{zz} = 1$. When we consider an active material that leads to gain in the FPC with DBE, we consider it diluted only in the isotropic layer of each unit cell in Fig. 1. The active isotropic materials is assumed to have a refractive index equal to $n = n' + in'' = 1 - i2.2 \times 10^{-4}$. The same imaginary part of $n$ is considered in one layer of the unit cell of the FPC with RBE, i.e., $n_2 = 1.5 - i2.2 \times 10^{-4}$ (keeping $n_1 = 3.2$ real). In the uniform FPC with gain in Fig. 1 we assume $n = 2.25 - i2.2 \times 10^{-4}$. All fitting curves report in this paper were performed by setting a root mean square error (RMSE) of 0.5%.

## APPENDIX B: CALCULATION OF THE LOCAL DENSITY OF STATES

We outline here the method we use to evaluate the local density of states through the dyadic Green's function in Cartesian coordinates for an electric sheet current in the *x-y* plane, and located at an arbitrary point $z'$ in a stack of anisotropic layered media shown in the inset of Fig. 1, and described by symmetric permittivity dyadic as in (A2), by generalizing the scheme used in [51,67,68] for stacks of isotropic layers. We consider a sheet current $\mathbf{J}(z') = J_0 \hat{\mathbf{p}} \delta(z - z')$ with $J_0$ being a constant ($J_0$ has units of [A/m]), that is only dependent on the coordinate $z$ and polarized along a unit vector $\hat{\mathbf{p}}$, where $\hat{\mathbf{p}}$ is strictly oriented in the *x-y* plane, i.e., $\hat{\mathbf{z}} \cdot \hat{\mathbf{p}} = 0$. Consequently, only transversely polarized electromagnetic plane waves (whose electric field is orthogonal to the *z*-directed wavevector) are excited due to such sheet current $\mathbf{J}(z')$. In other words, the direction of energy flux of those propagating waves inside the stack is parallel to their wavevector, both along the *z*-direction. For these reasons we conveniently construct the transverse electric field response function to the Dirac delta sheet current $\mathbf{J}(z')$, which we identify as the 2×2 matrix (dyadic) Green's function. First, the transverse electric field satisfies the inhomogeneous one-dimensional vector wave equation [76,77]

$$\left( \underline{\underline{\mathbf{1}}} \frac{\partial^2}{\partial z^2} + \underline{\underline{\boldsymbol{\varepsilon}}}_t(z) k_0^2 \right) \mathbf{E}(z, z') = -i\omega\mu_0 \mathbf{J}(z'), \quad (B1)$$

where $\underline{\underline{\mathbf{1}}}$ is a 2×2 unit dyadic. Accordingly, 2×2 matrix (dyadic) Green's function $\underline{\underline{\mathbf{G}}}(z, z')$ is constructed by solving the following equation

$$\left(\underline{\underline{\mathbf{1}}}\frac{\partial^2}{\partial z^2}+k_0^2\underline{\underline{\boldsymbol{\varepsilon}}}_t(z)\right)\underline{\underline{\mathbf{G}}}(z,z')=-\delta(z-z')\underline{\underline{\mathbf{1}}}, \quad (B2)$$

with the appropriate boundary conditions [42,43], i.e., continuity of $\underline{\underline{\mathbf{G}}}(z,z')$ for all values of $z$, the jump discontinuity in its first derivative in $z$ due to a Dirac delta sheet current, as well as the outgoing waves condition [51,67,68] outside the cavity in both the $\pm z$ direction. When considering a Cartesian reference system, Green's function is a 2×2 matrix of the form

$$\underline{\underline{\mathbf{G}}}(z,z')=\begin{pmatrix}G_{xx}(z,z') & G_{xy}(z,z') \\ G_{yx}(z,z') & G_{yy}(z,z')\end{pmatrix}. \quad (B3)$$

Detailed studies of Green's functions, their properties and construction methods can be found in many mathematical physics textbooks such as Refs. [78–80], and dyadic formulations can be found in [60,76,77,81]. For our purposes we only calculate the relevant parts of the dyadic Green's function which contribute to the LDOS, as follows.

We define a projected (or partial) local density of states [60,62,63,66] pertaining to an arbitrary transverse-to-$z$ source orientation denoted by a unit vector $\hat{\mathbf{p}}(\phi)=\cos\phi\hat{\mathbf{x}}+\sin\phi\hat{\mathbf{y}}$ where $\phi$ spans the angular range from 0 to $2\pi$. This projected LDOS is proportional to the imaginary part of the Green's function and given by [63,66]

$$\rho(z,\hat{\mathbf{p}})=-\frac{2k_0}{\pi c}\left(\mathrm{Im}\left(\hat{\mathbf{p}}\cdot\left(\underline{\underline{\mathbf{G}}}(z,z)\cdot\hat{\mathbf{p}}\right)\right)\right). \quad (B4)$$

The average of the projected LDOS over all possible orientation, as obtained in (3) is calculated as

$$\rho(z)=\langle\rho(z,\hat{\mathbf{p}})\rangle_{\hat{\mathbf{p}}}$$
$$=-\frac{k_0}{\pi^2 c}\int_0^{2\pi}\mathrm{Im}\left(\hat{\mathbf{p}}(\phi)\cdot\left(\underline{\underline{\mathbf{G}}}(z,z)\cdot\hat{\mathbf{p}}(\phi)\right)\right)d\phi$$
$$=-\frac{k_0}{\pi c}\mathrm{Im}\left(\mathrm{Tr}\left[\underline{\underline{\mathbf{G}}}(z,z)\right]\right), \quad (B5)$$

where Tr denotes the trace. Therefore, the averaged LDOS in (3) is proportional to the trace of the imaginary part of $\underline{\underline{\mathbf{G}}}(z,z)$ [60,63]. Therefore, instead of calculating the full dyadic Green's function, for our purpose we need to calculate only the term $\mathrm{Im}(G_{xx}(z,z)+G_{yy}(z,z))$, utilizing the Poynting theorem [60,76]. For this purpose we assume a unit current sheet oriented along the $x$-direction, whose current density takes the form $\mathbf{J}=\delta(z-z')\hat{\mathbf{x}}$, for instance. The time-average power density (per-unit-area) emitted by such current sheet according to the Poynting theorem [76,77] is given by

$$P(z,\hat{\mathbf{x}})=-\frac{1}{2}\mathrm{Re}\left(\int\mathbf{J}^*\cdot\mathbf{E}(z',\hat{\mathbf{x}})dz'\right)$$
$$=-\frac{1}{2}\mathrm{Re}(E_x(z,\hat{\mathbf{x}})), \quad (B6)$$

where $\mathrm{Re}(E_x(z,\hat{\mathbf{x}}))$ is real part of the $x$-polarized electric field at point $z$ due to a unit current sheet along the $x$-direction at the same point $z$. This field component is proportional to a term of dyadic Green's function in (B3) as $\mathrm{Im}(G_{xx}(z,z))=\omega\mu_0\mathrm{Re}(E_x(z,\hat{\mathbf{x}}))$ [76,77], implying that

$$\mathrm{Im}(G_{xx}(z,z))=\frac{-2}{\omega\mu_0}P(z,\hat{\mathbf{x}}). \quad (B7)$$

Hence calculating the power density (the right hand side of (B7)) emitted by the current sheet using the transfer matrix method allows us to compute the imaginary part of $G_{xx}(z,z)$. Similarly, $\mathrm{Im}(G_{yy}(z,z))$ is obtained by evaluating the power density emitted by a $y$-directed unit current sheet. The emitted power density for each of the two polarized current sources is easily calculated using the transfer matrix approach for the layered media [30–33], following Appendix A. An important consequence of (B7) is the intimate relation between the LDOS and the power emitted by the source [51,60,82]. From both (B7) and (B5) one can write the LDOS enhancement factor, which is plotted in Fig. 4(a-c), in the following form

$$\frac{\rho(z)}{\rho_{\mathrm{free\,space}}}=\frac{P(z)}{P_{\mathrm{free\,space}}}$$
$$=-k_0\mathrm{Im}(G_{xx}(z,z)+G_{yy}(z,z)), \quad (B8)$$

where $P(z)$ is the emitted power density (per-unit-area) averaged over all source orientations inside the layered media.

**REFERENCES**

[1] K. Inoue, M. Sasada, J. Kawamata, K. Sakoda, and J. W. Haus, Jpn. J. Appl. Phys. **38**, L157 (1999).
[2] H.-Y. Ryu, S.-H. Kwon, Y.-J. Lee, Y.-H. Lee, and J.-S. Kim, Appl. Phys. Lett. **80**, 3476 (2002).
[3] J. M. Bendickson, J. P. Dowling, and M. Scalora, Phys. Rev. E **53**, 4107 (1996).

[4] B.-S. Song, S. Noda, T. Asano, and Y. Akahane, Nat. Mater. **4**, 207 (2005).
[5] B. Maune, M. Lončar, J. Witzens, M. Hochberg, T. Baehr-Jones, D. Psaltis, A. Scherer, and Y. Qiu, Appl. Phys. Lett. **85**, 360 (2004).
[6] F. Scotognella, D. P. Puzzo, A. Monguzzi, D. S. Wiersma, D. Maschke, R. Tubino, and G. A. Ozin, Small **5**, 2048 (2009).
[7] A. Figotin and I. Vitebskiy, Phys. Rev. B **67**, 165210 (2003).
[8] F. D. M. Haldane and S. Raghu, Phys. Rev. Lett. **100**, 013904 (2008).
[9] H. Ramezani, S. Kalish, I. Vitebskiy, and T. Kottos, Phys. Rev. Lett. **112**, 043904 (2014).
[10] N. V. Bloch, K. Shemer, A. Shapira, R. Shiloh, I. Juwiler, and A. Arie, Phys. Rev. Lett. **108**, 233902 (2012).
[11] I. Bayn, B. Meyler, A. Lahav, J. Salzman, R. Kalish, B. A. Fairchild, S. Prawer, M. Barth, O. Benson, and T. Wolf, Diam. Relat. Mater. **20**, 937 (2011).
[12] S. Noda, M. Fujita, and T. Asano, Nat. Photonics **1**, 449 (2007).
[13] I. L. Garanovich, S. Longhi, A. A. Sukhorukov, and Y. S. Kivshar, Phys. Rep.-Rev. Sect. Phys. Lett. **518**, 1 (2012).
[14] J. D. Joannopoulos, S. G. Johnson, J. N. Winn, and R. D. Meade, *Photonic Crystals: Molding the Flow of Light* (Princeton university press, 2011).
[15] C. M. Soukoulis, *Photonic Crystals and Light Localization in the 21st Century* (Springer Science & Business Media, 2012).
[16] S. K. Selvaraja, P. Jaenen, W. Bogaerts, D. Van Thourhout, P. Dumon, and R. Baets, Light. Technol. J. Of **27**, 4076 (2009).
[17] P. Yao, V. S. C. Manga Rao, and S. Hughes, Laser Photonics Rev. **4**, 499 (2010).
[18] C. Sciancalepore, B. B. Bakir, X. Letartre, J. Harduin, N. Olivier, C. Seassal, J.-M. Fedeli, and P. Viktorovitch, Photonics Technol. Lett. IEEE **24**, 455 (2012).
[19] D.-H. Ko, J. R. Tumbleston, L. Zhang, S. Williams, J. M. DeSimone, R. Lopez, and E. T. Samulski, Nano Lett. **9**, 2742 (2009).
[20] G. Demésy and S. John, J. Appl. Phys. **112**, 074326 (2012).
[21] S. Eyderman, A. Deinega, and S. John, J. Mater. Chem. A **2**, 761 (2014).
[22] A. Chutinan and S. John, Phys. Rev. A **78**, 023825 (2008).
[23] A. Goban, C.-L. Hung, S.-P. Yu, J. D. Hood, J. A. Muniz, J. H. Lee, M. J. Martin, A. C. McClung, K. S. Choi, and D. E. Chang, Nat. Commun. **5**, (2014).
[24] S. John and T. Quang, Phys. Rev. A **50**, 1764 (1994).
[25] U. Hoeppe, C. Wolff, J. Küchenmeister, J. Niegemann, M. Drescher, H. Benner, and K. Busch, Phys. Rev. Lett. **108**, 043603 (2012).
[26] M. Povinelli, S. Johnson, and J. Joannopoulos, Opt. Express **13**, 7145 (2005).
[27] S. Kubo, D. Mori, and T. Baba, Opt. Lett. **32**, 2981 (2007).
[28] S. Noda and T. Baba, *Roadmap on Photonic Crystals* (Springer Science & Business Media, 2013).
[29] A. Figotin and I. Vitebskiy, Phys. Rev. E **68**, 036609 (2003).
[30] A. Figotin and I. Vitebskiy, Phys. Rev. E **72**, 036619 (2005).
[31] A. Figotin and I. Vitebskiy, Phys. Rev. E **74**, 066613 (2006).
[32] A. Figotin and I. Vitebskiy, Phys. Rev. A **76**, 053839 (2007).
[33] A. Figotin and I. Vitebskiy, Laser Photonics Rev. **5**, 201 (2011).
[34] J. P. Dowling, M. Scalora, M. J. Bloemer, and C. M. Bowden, J. Appl. Phys. **75**, 1896 (1994).
[35] R. Colombelli, K. Srinivasan, M. Troccoli, O. Painter, C. F. Gmachl, D. M. Tennant, A. M. Sergent, D. L. Sivco, A. Y. Cho, and F. Capasso, Science **302**, 1374 (2003).
[36] A. E. Siegman, *Lasers. Mill Valley* (University Science Books, 1986), pg.102-108.
[37] J. T. Verdeyen, Laser Electron. Ed. JT Verdeyen Englewood Cliffs NJ Prentice Hall 1989 640 P **1**, (1989).
[38] K. F. Renk, *Basics of Laser Physics: For Students of Science and Engineering* (Springer Science & Business Media, 2012), pg. 33-39.
[39] Y. Amnon, *Quantum Electronics*, 3rd Edition (Wiley, 1989).
[40] L. A. Coldren, S. W. Corzine, and M. L. Mashanovitch, *Diode Lasers and Photonic Integrated Circuits* (John Wiley & Sons, 2012), pg. 220-237.
[41] O. Boyraz and B. Jalali, Opt. Express **12**, 5269 (2004).
[42] P. M. Becker, A. A. Olsson, and J. R. Simpson, *Erbium-Doped Fiber Amplifiers: Fundamentals and Technology* (Academic press, 1999), pg. 167, 179.
[43] P. Yeh and C. Gu, *Optics of Liquid Crystal Displays* (John Wiley & Sons, 2010).
[44] J. R. Burr, N. Gutman, C. Martijn de Sterke, I. Vitebskiy, and R. M. Reano, Opt. Express **21**, 8736 (2013).
[45] N. Gutman, C. Martijn de Sterke, A. A. Sukhorukov, and L. C. Botten, Phys. Rev. A **85**, (2012).
[46] N. Gutman, W. H. Dupree, Y. Sun, A. A. Sukhorukov, and C. M. de Sterke, Opt. Express **20**, 3519 (2012).
[47] M. G. Wood, J. R. Burr, and R. M. Reano, Opt. Lett. **40**, 2493 (2015).
[48] A. Figotin and I. Vitebskiy, ArXiv09091393 Phys. (2009).
[49] J. Grgić, J. R. Ott, F. Wang, O. Sigmund, A.-P. Jauho, J. Mørk, and N. A. Mortensen, Phys. Rev. Lett. **108**, 183903 (2012).
[50] M. Fox, *Quantum Optics: An Introduction: An Introduction* (Oxford University Press, 2006), pg. 167-204.
[51] G. D'Aguanno, N. Mattiucci, M. Scalora, M. J. Bloemer, and A. M. Zheltikov, Phys. Rev. E **70**, 016612 (2004).
[52] A. Yariv and P. Yeh, *Photonics: Optical Electronics in Modern Communications (The Oxford Series in Electrical and Computer Engineering)* (Oxford University Press, Inc., 2006).
[53] F. T. Smith, Phys. Rev. **118**, 349 (1960).
[54] G. L. Matthaei, L. Young, and E. M. Jones, *Design of Microwave Filters, Impedance-Matching Networks, and Coupling Structures. Volume 2* (DTIC Document, 1963), pg. 339-341.
[55] H. G. Winful, New J. Phys. **8**, 101 (2006).
[56] T. Laupêtre, C. Proux, R. Ghosh, S. Schwartz, F. Goldfarb, and F. Bretenaker, Opt. Lett. **36**, 1551 (2011).
[57] H.-Y. Yao, N.-C. Chen, T.-H. Chang, and H. G. Winful, Phys. Rev. A **86**, 053832 (2012).
[58] G. F. Franklin, J. D. Powell, and A. Emami-Naeini, Pretince Hall Inc (2006).
[59] T. Quang, M. Woldeyohannes, S. John, and G. S. Agarwal, Phys. Rev. Lett. **79**, 5238 (1997).
[60] L. Novotny and B. Hecht, *Principles of Nano-Optics* (Cambridge university press, 2012).
[61] H. Yokoyama and K. Ujihara, *Spontaneous Emission and Laser Oscillation in Microcavities* (CRC press, 1995), pg. 15-42.
[62] A. A. Asatryan, K. Busch, R. C. McPhedran, L. C. Botten, C. Martijn de Sterke, and N. A. Nicorovici, Phys. Rev. E **63**, 046612 (2001).
[63] D. P. Fussell, R. C. McPhedran, and C. M. De Sterke, Phys. Rev. E **70**, 066608 (2004).
[64] J. A. E. Wasey and W. L. Barnes, J. Mod. Opt. **47**, 725 (2000).
[65] A. F. Koenderink, M. Kafesaki, C. M. Soukoulis, and V. Sandoghdar, JOSA B **23**, 1196 (2006).
[66] K. Joulain, R. Carminati, J.-P. Mulet, and J.-J. Greffet, Phys. Rev. B **68**, 245405 (2003).
[67] O. D. Stefano, S. Savasta, and R. Girlanda, J. Mod. Opt. **48**, 67 (2001).
[68] G. D'Aguanno, M. Centini, M. Scalora, C. Sibilia, M. Bertolotti, M. J. Bloemer, and C. M. Bowden, JOSA B **19**, 2111 (2002).
[69] S. John and T. Quang, Phys. Rev. Lett. **74**, 3419 (1995).
[70] N. Vats and S. John, Phys. Rev. A **58**, 4168 (1998).
[71] J. Grgić, E. Campaioli, S. Raza, P. Bassi, and N. A. Mortensen, Opt. Quantum Electron. **42**, 511 (2011).
[72] E. Yeganegi, A. Lagendijk, A. P. Mosk, and W. L. Vos, Phys. Rev. B **89**, 045123 (2014).
[73] S. John and J. Wang, Phys. Rev. Lett. **64**, 2418 (1990).
[74] K.-Y. Jung and F. L. Teixeira, Phys. Rev. B **77**, 125108 (2008).
[75] V. A. Tamma, A. Figotin, and F. Capolino, ArXiv13107645 Phys. (2013).
[76] R. E. Collin, Field Theory of Guided Waves, (Wiley-Interscience 1960), pg. 56-103.

[77] J. G. Van Bladel, *Electromagnetic Fields* (John Wiley & Sons, 2007), pg. 11-17 and pg. 1053-1045.
[78] P. Dennery and A. Krzywicki, *Mathematics for Physicists* (Courier Dover Publications, 1996), pg. 273-299.
[79] B. Friedman, *Principles and Techniques of Applied Mathematics* (New York, 1961).
[80] R. Courant and D. Hilbert, *Methods of Mathematical Physics* (CUP Archive, 1966).
[81] C.-T. Tai, *Dyadic Green Functions in Electromagnetic Theory* (IEEE press New York, 1994).
[82] X. Liang and S. G. Johnson, Opt. Express **21**, 30812 (2013).